\documentclass[12pt]{article}

\usepackage[utf8]{inputenc}      
\usepackage[T1]{fontenc}         
\usepackage{lmodern}             
\usepackage[margin=1in]{geometry} 
\usepackage{setspace}                      
\usepackage{times}               
\usepackage{hyperref}            
\hypersetup{
    colorlinks = true,
    linkcolor  = blue,
    citecolor  = blue,
    urlcolor   = blue
}
\usepackage{amsmath,amssymb,amsfonts,amsthm} 
\usepackage{mathtools}                      

\usepackage{graphicx}   
\usepackage{subcaption} 
\usepackage{float}      
\usepackage{booktabs}   
\usepackage{multirow}   
\usepackage{caption}
\usepackage{tabularx}
\usepackage{threeparttable}
\usepackage{makecell} 

\usepackage{natbib}

\usepackage{xcolor}        
\usepackage{listings}      
\usepackage{enumitem}      
\usepackage{authblk}
\usepackage{pdfpages}

\newcolumntype{Y}{>{\centering\arraybackslash}X}

\title{Scalable Foundation Interatomic Potentials via Message-Passing Pruning and Graph Partitioning}

\author{Lingyu Kong$^\dag$, Jaeheon Shim$^\dag$, Guoxiang Hu$^\ddag$, Victor Fung$^\dag$ \\
$^\dag$School of Computational Science and Engineering, Georgia Institute of Technology\\
$^\ddag$School of Materials Science and Engineering, Georgia Institute of Technology\\
Corresponding author: \texttt{victorfung@gatech.edu} \\
}

\date{}

\begin{document}

\maketitle

\begin{abstract}
    Atomistic foundation models (AFMs) have great promise as accurate interatomic potentials, and have enabled data-efficient molecular dynamics simulations with near quantum mechanical accuracy. However, AFMs remain markedly slower at inference and are far more memory-intensive than conventional interatomic potentials, due to the need to capture a wide range of chemical and structural motifs in pre-training datasets requiring deep, parameter-rich model architectures. These deficiencies currently limit the practical use of AFMs in molecular dynamics (MD) simulations at extended temporal and spatial scales. To address this problem, we propose a general workflow for accelerating and scaling AFMs containing message-passing architectures. We find that removing low-contribution message-passing layers from AFM backbones serves as an effective pruning method, significantly reducing the parameter count while preserving the accuracy and data-efficiency of AFMs. Once pruned, these models become more accessible for large scale simulations via a graph-partitioned, GPU-distributed strategy, which we implement and demonstrate within the AFM fine-tuning platform MatterTune. We show that this approach supports million-atom simulations on both single and multiple GPUs, and enables task-specific large-scale simulations at nanosecond timescales with AFM-level accuracy.  
\end{abstract}

\section{Introduction}

Atomistic molecular dynamics (MD) simulations capture the evolution of atomic configurations over time and allow for the mapping of microscopic interactions onto macroscopic observables, making them indispensable for uncovering reaction mechanisms, predicting material behavior, and guiding the design and screening of new materials ~\cite{cheng2017full, jiabo2021modeling, cheng2019ab, mo2012first, islamov2023high, richards2016design, mo2014insights}. A central challenge in computational materials science lies in extending the spatial and temporal limits of these atomistic simulations, thereby enabling the study of increasingly complex phenomena such as phase transitions, defect migration, or chemical reactions in realistic environments. Achieving such scales requires models that can evaluate the potential energy surface (PES) of a given system with both high efficiency and fidelity.

Ab initio methods, such as density functional theory (DFT) and coupled cluster (CC) approaches~\cite{hohenberg1964inhomogeneous, kohn1965self, raghavachari1989fifth}, , are capable of highly accurate evaluation of the PES, providing high-fidelity results with no parameterization needed. However, the steep computational scaling of these methods with respect to system size severely limits their practical applicability, restricting simulations to relatively short timescales and small system sizes. While they remain the gold standard for benchmarking and for training parameterized models, their direct use in large-scale or long-timescale molecular dynamics remains computationally prohibitive. This limitation has motivated the search for alternative strategies that retain near-ab initio accuracy while significantly reducing computational cost. ~\cite{beck2000real, payne1992iterative, kresse1996efficiency, marx2000ab, tuckerman2002ab}.

Machine learning interatomic potentials (MLIPs) ~\cite{bartok2010gaussian, schutt2017schnet, schutt2021equivariant, wang2018deepmd, drautz2019atomic, chen2019graph, chen2022universal, batzner20223, batatia2022mace, musaelian2023learning, liao2022equiformer} are a rapidly growing class of methods that overcome this computational bottleneck by serving as an efficient alternative to first-principles calculations. Through the training of parameterized models, MLIPs can reduce the evaluation cost of the PES to linear or even sublinear scaling, accelerating PES predictions by several orders of magnitude while maintaining near-quantum accuracy. Among these, graph neural networks (GNNs) are a widely used architecture in MLIP models, with the capability to effectively learn long-range and high-order atomic interactions in a data-driven manner, particularly for structurally diverse and compositionally complex systems. The use of these MLIP models to accelerate MD simulations has been extensively studied, and mature interfaces with MD simulation packages have been established in practice ~\cite{wang2018deepmd, thompson2022lammps, schütt2018schnetpack, drautz2019atomic, novikov2020mlip, kapil2019pi}. 

Recently, the concept of Atomistic Foundation Models (AFMs) has been proposed ~\cite{batatia2023foundation, merchant2023scaling, zhang2024dpa, shoghi2023molecules, yang2024mattersim, neumann2024orbfastscalableneural, rhodes2025orbv3atomisticsimulationscale, liao2023equiformerv2, fu2025learning} as a class of models with broad or "universal" applicability across the periodic table and with comparable or sometimes better accuracy than bespoke MLIPs. These models are constructed via pre-training on large, diverse atomistic datasets, learning high-quality, general-purpose material embeddings that transfer across a broad chemical and structural space. Early studies ~\cite{yang2024mattersim, kong2025mattertune, radova2025fine, kaur2025data} show that AFMs can be further fine-tuned on only a handful of task-specific data and still match or even surpass the bespoke MLIPs in terms of accuracy, significantly lowering training costs for MLIPs and greatly enabling widespread use of these models in simulations.

However, despite their superior accuracy and data-efficiency, AFMs remain difficult to deploy in production-level MD simulations, especially those spanning nanoseconds and involving more than tens of thousands of atoms ~\cite{du2023machine, jumper2021highly, shi2020high, holstun2025accelerating}. To capture the complex chemical and geometric relationships present in large, element-wide training sets, AFMs generally rely on sophisticated, high-capacity architectures to maximize expressivity; for instance many message-passing layers are introduced in the models to better capture medium- and long-range interactions, further inflating model size and complexity. For example, the MatterSim V1 ~\cite{yang2024mattersim} model employs 3-4 message-passing layers, JMP ~\cite{shoghi2023molecules} uses 4-6 layers, and ORB V2 ~\cite{neumann2024orbfastscalableneural} incorporates up to 15 message-passing layers. These design choices lead to two practical bottlenecks: (i) AFMs run markedly slower than conventional MLIPs and cannot keep pace with the millions of timesteps required for nanosecond-scale MD;
(ii) their much larger memory footprint at inference restricts the maximum structure size that can be processed on a single GPU. 

To overcome the slow inference and high memory demand of AFMs and thereby position them as capable replacements for conventional MLIPs in MD simulations, we propose an acceleration workflow that satisfies two general requirements: (i) that it should be reasonably architecture agnostic, applying broadly to most AFMs rather than being custom-tailored to a single architecture; and (ii) that it should impose only minimal modifications to existing AFM implementations, ensuring flexible and straightforward adoption. 

Our approach is inspired by the over-smoothing phenomenon observed in graph neural networks: as message passing grows deeper, node features tend to homogenize and the incremental information gained at each layer diminishes ~\cite{li2018deeper, oono2019graph, cai2020note, chen2020measuring}. Motivated by these insights, we compared the node embeddings produced at every message-passing layer in three representative AFMs: MatterSim, JMP, and ORB, and verified that layers at different depths contribute unevenly to the final representation. The early layers provide most of the useful signal, whereas the deepest layers add relatively little. Based on this finding, we designed a pruning scheme that keeps the first $K$ message-passing layers and removes the other low-contribution ones, yielding a lightweight version of an AFM. Fine-tuning experiments and accuracy analyses show that the pruned AFMs, even when only one or two layers are retained, preserve the accuracy and data-efficiency superiority of AFMs. 

This pruning strategy, which we term Message-Passing Depth Pruning (MPDP), not only reduces the parameter count of AFMs and thus their inference costs, it also makes the model better suited to the second stage of our proposed acceleration workflow: graph-partitioned multi-GPU parallelism. A central challenge in enabling multi-GPU parallelization of potential models via graph partitioning lies in preserving the effective receptive field, which is the spatial extent over which information can be aggregated, of every atom. For a model with local cutoff $r_c$ and $M$ layers of message passing, the effective receptive field cutoff is consequently extended to become $r_{\text{eff}}=M\times r_c$. One approach to parallelization relies on shared buffers or explicit inter-GPU communication to propagate node messages across devices, thereby maintaining global information flow. This does, however, introduce a communication overhead that grows approximately linearly with the number of message-passing layers, which can be costly on hardware without high-bandwidth interconnects such as NVLink. An alternative approach utilizes the so-called halo method, where each subgraph is pre-expanded to contain all neighbors inside the receptive field at the time of partitioning. However, as the effective receptive field expands roughly cubically with the message-passing depth $M$, this strategy can lead to substantial computational and memory overhead. However, by pruning the layers of the model with the MPDP approach, we can reduce the original depth $M$---which is typically large---to a much smaller value $K$. This substantially alleviates the overhead associated with the halo-based graph partitioning scheme, thereby significantly improving its efficiency with a small loss in accuracy. Beyond this, the halo method also offers several additional advantages: it completely eliminates the need for inter-GPU communication during inference, and its implementation is straightforward without requiring any modifications to the underlying model architecture. Taken together, these properties make the halo-based parallelization scheme particularly well-suited to our pruned models, and we adopt this strategy in the present work.

By integrating the MPDP method with graph-partitioned multi-GPU parallelism, we have designed an acceleration workflow applicable to any AFM that relies on deep message passing. The procedure first applies MPDP, retaining only the first $K$ message-passing layers, then fine-tunes the pruned model, and finally partitions the graph to either run on a single GPU in serial, or distribute inference across multiple GPUs for an additional speedup. It is worth noting that a number of recent studies have explored multi-GPU acceleration of MLIPs and AFMs. \texttt{chemtrain-deploy}~\cite{fuchs2025chemtrain} adopts the aforementioned halo-based spatial domain decomposition of LAMMPS with a focus on JAX-defined ~\cite{bradbury2018jax} models. Meanwhile, DistMLIP~\cite{han2025distmlip} employs direct atomic graph partitioning with explicit inter-GPU communication at each message-passing layer. Our approach differs in the inclusion of the Message-Passing Depth Pruning (MPDP) strategy, which substantially reduces model depth and synergizes well with the halo-based partitioning strategy, enabling efficient and communication-free parallelization. We furthermore integrate this parallelism with AFM finetuning to 
enable nanosecond-scale MD simulations with AFM-level quality of systems containing tens of thousands or more atoms, making the previously cumbersome AFMs powerful engines for realistic large-scale MD.

\begin{figure}[H]
  \vspace{-0.4cm}
  \centering
  \includegraphics[width=0.9\linewidth]{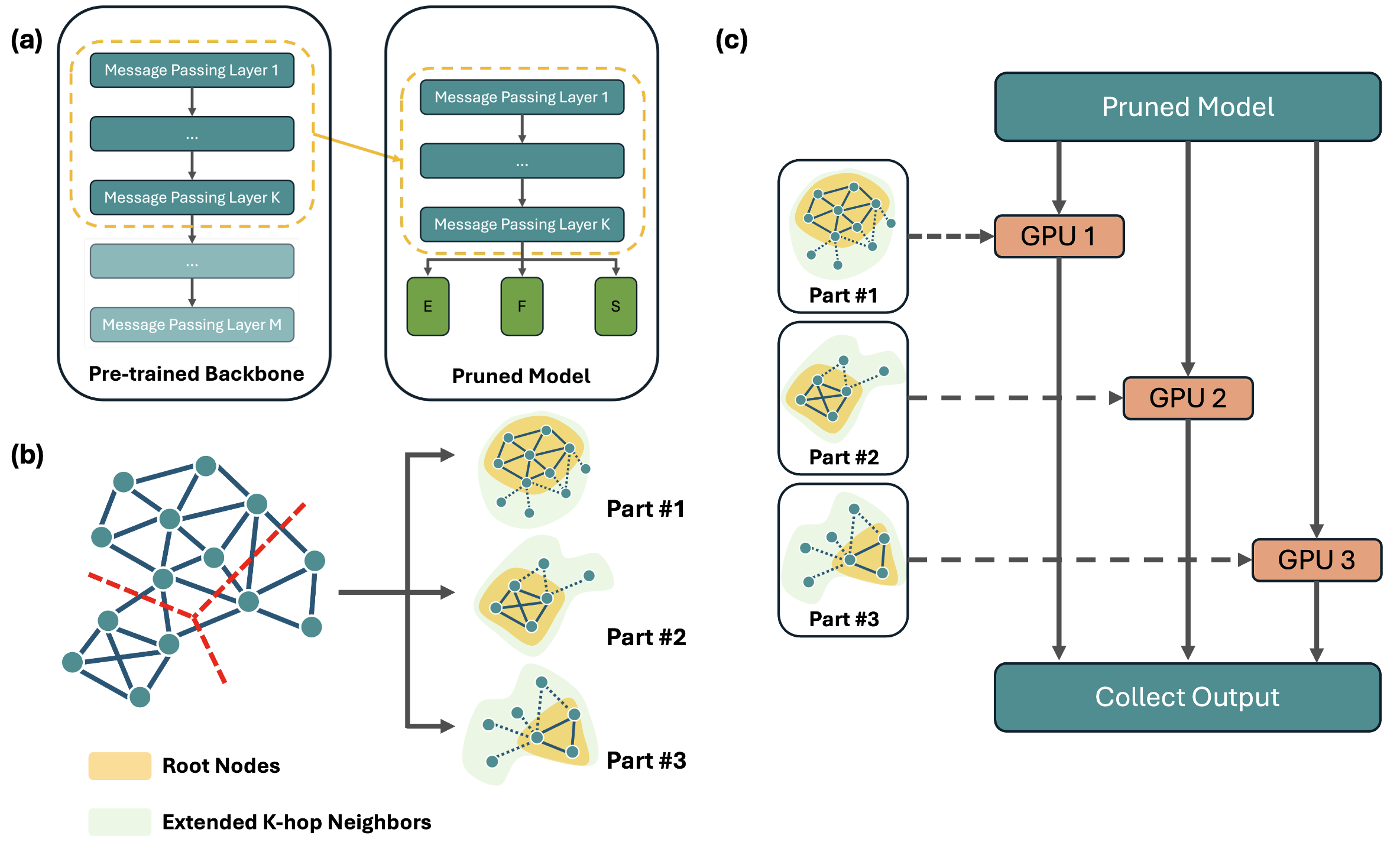}
  \caption{\textbf{Overview of the pruning‑and‑partition acceleration workflow for AFMs.} \textbf{(a) Message-passing depth pruning.} A pre‑trained AFM backbone with M message‑passing layers is truncated to the first K layers (K$<$M). The node and edge features after layer K feed directly into output heads for energy, forces, and stress. \textbf{(b) Graph partitioning.} The atomic graph is split by red dashed lines; the nodes in each piece define the set of root nodes. To preserve the receptive field of the pruned model, each piece is expanded to include the K-hop neighbors of its root nodes, following the original adjacency, yielding extended subgraphs. \textbf{(c) Multi‑GPU inference.} The extended subgraphs are assigned to different GPUs for independent forward passes, and their outputs are gathered to produce predictions for the full structure.}
  \label{fig:overview}
  \vspace{-0.4cm}
\end{figure}

\section{Results}

In the following, we demonstrate the utility of the proposed method in three illustrative case studies: (i) simulating quenching of molten Li\textsubscript{3}PO\textsubscript{4} phosphate electrode, (ii) modeling phase segregation during the melt–quench process of the mixed silicon-silica system, and (iii) exploring metal-support interactions on a oxide-supported metal catalyst. These simulations span from tens of thousands to approximately half a million atoms with trajectory lengths ranging from 50 ps to several nanoseconds. With pruned, partition-parallel AFMs in place, all simulations can be completed within practical wall-clock budgets while yielding physically consistent structural and dynamical behavior.
\subsection{Melt–Quench Formation of Amorphous Li\textsubscript{3}PO\textsubscript{4} and Li-Ion Diffusion}

Li\textsubscript{3}PO\textsubscript{4} is widely investigated as a solid‐state electrolyte due to its high Li-ion conductivity and electrochemical stability making it a promising replacement for liquid electrolytes in Li-ion batteries ~\cite{yu1997stable, bates2000thin, westover2020plasma, li2017study}. A prior work by Musaelian et al. ~\cite{musaelian2023learning} demonstrated the training of a lightweight Allegro model on a provided dataset containing Li\textsubscript{3}PO\textsubscript{4} melting and quenching trajectories, which serves as a representative benchmark for speed and scalability in massively parallel MD simulations. 

Following the same dataset splitting method in \cite{musaelian2023learning}, we sampled 5\,000 frames from melting and quenching trajectory separately to construct a 10\,000‑structure training set (hereafter denoted 10k‑sample). To highlight the superior data efficiency of AFMs, we additionally constructed another ultra‑small training set by selecting only the first 100 frames from the melting trajectory (denoted 100‑sample). We fine‑tuned the AFMs MatterSim, MACE, JMP, and ORB onboth the 10k‑sample and 100‑sample datasets and report the results in Table~\ref{table:li3po4-mae}. From these results we observe that, for a given AFM backbone, fine‑tuning on the 10k‑sample versus the 100‑sample dataset yields only modest performance differences. This highlights a key advantage of AFMs over conventional MLIPs, namely their ability to leverage pretrained representations to transfer across chemical and structural variations without relying on large training datasets. Furthermore, most AFM variants fine‑tuned on as few as 100 samples can already match or surpass the accuracy of the Allegro baseline trained on the full 10k‑sample set and, of particular note, the 100-sample dataset consists only the first 100 frames of the melting trajectory and contains no quenching structures. This demonstrates that, with virtually no prior coverage of the target configuration space, it is possible to obtain a model of competitive accuracy using only a minimal amount of training data. In conventional AFMs, this comes with a significant trade-off in much slower simulation speeds and size constraints. We subsequently show that these limitations can be overcome through graph partitioning and parallelization.

\begin{table}[h]
  \centering
  \fontsize{8pt}{11pt}\selectfont
  \begin{threeparttable}
    \caption{\textbf{Fine-tuning performance of various AFMs on Li\textsubscript{3}PO\textsubscript{4} and comparison with the Allegro baseline}}
    \label{table:li3po4-mae}
    \begin{tabularx}{0.9\linewidth}{l l| *{2}{Y}| *{2}{Y}}
      \toprule
      &  & \multicolumn{2}{c|}{\textbf{10k-sample}} 
         & \multicolumn{2}{c}{\textbf{100-sample}} \\
      \cmidrule(lr){3-4}\cmidrule(lr){5-6}
      &  & $\mathrm{MAE}_{E}$ (meV/atom) 
         & $\mathrm{MAE}_{F}$ (meV/\AA)
         & $\mathrm{MAE}_{E}$ (meV/atom) 
         & $\mathrm{MAE}_{F}$ (meV/\AA) \\
      \midrule
      \textbf{Allegro} & & 1.7 & 73.40 & -- & -- \\
      \midrule
      \multirow{3}{*}{\makecell[l]{\textbf{MatterSim\ss{}}\\\textbf{V1-1M}}}
        & MP$\times$1 & 0.6 & 56.83 & 8.5 & 73.17 \\
        & MP$\times$2 & 0.5 & 37.93 & 10.6 & 60.17 \\
        & MP$\times$3\textsuperscript{*} & 0.3 & 32.60 & 4.3 & 50.57 \\
      \midrule
      \multirow{3}{*}{\textbf{JMP-S}}
        & MP$\times$1 & 0.6 & 21.07 & 111.6 & 73.26 \\
        & MP$\times$2 & 0.6 & 14.32 & 120.1 & 38.68 \\
        & MP$\times$4\textsuperscript{*} & 0.4 & 11.17 & 84.2 & 29.78 \\
      \midrule
      \multirow{3}{*}{\textbf{ORB-V3-Omat}}
        & MP$\times$1 & 0.3 & 32.60 & 137.01 & 259.50 \\
        & MP$\times$3 & 1.2 & 24.94 & 9.3 & 68.05 \\
        & MP$\times$5\textsuperscript{*} & 1.0 & 19.98 & 11.2 & 73.38 \\
      \midrule
      \multirow{2}{*}{\textbf{MACE-M-Omat}}
        & MP$\times$1 & 0.46 & 31.76 & 53.45 & 51.87 \\
        & MP$\times$2\textsuperscript{*} & 0.19 & 20.96 & 14.94 & 25.86 \\
      \bottomrule
    \end{tabularx}

    \begin{tablenotes}[flushleft]
      \footnotesize
      \item The Allegro baseline MAEs are taken from~\cite{musaelian2023learning}.  
            MP$\times K$ denotes a pruned model retaining the first $K$ message‑passing layers. An asterisk superscript MP$\times M$\textsuperscript{*} indicates that the original backbone natively has $M$ message‑passing layers; i.e., the model is unpruned. The designation 10k‑sample refers to models trained on the complete AIMD dataset containing 10 000 structures, whereas 100‑sample denotes models trained solely on the first 100 frames of the melting trajectory. 
    \end{tablenotes}
  \end{threeparttable}
\end{table}

We then applied varying degrees of MPDP to each AFM, denoting a variant that retains only the first $K$ message-passing layers as MP$\times K$. The pruned models were also fine-tuned on both datasets, with results also summarized in Table~\ref{table:li3po4-mae}. Remarkably, we found that, in most cases, pruning the message-passing stack lead to an unacceptable loss in accuracy; even in the extreme case where only a single message-passing layer is retained, AFMs fine-tuned on the 100-sample dataset can achieve a performance comparable to the Allegro baseline. These results partially substantiate our hypothesis that AFMs suffer from over‑smoothing and contain deep message‑passing layers with marginal contribution, thus supporting the proposed pruning strategy. 

\begin{figure}[H]
  \vspace{-0.4cm}
  \centering
  \includegraphics[width=0.9\linewidth]{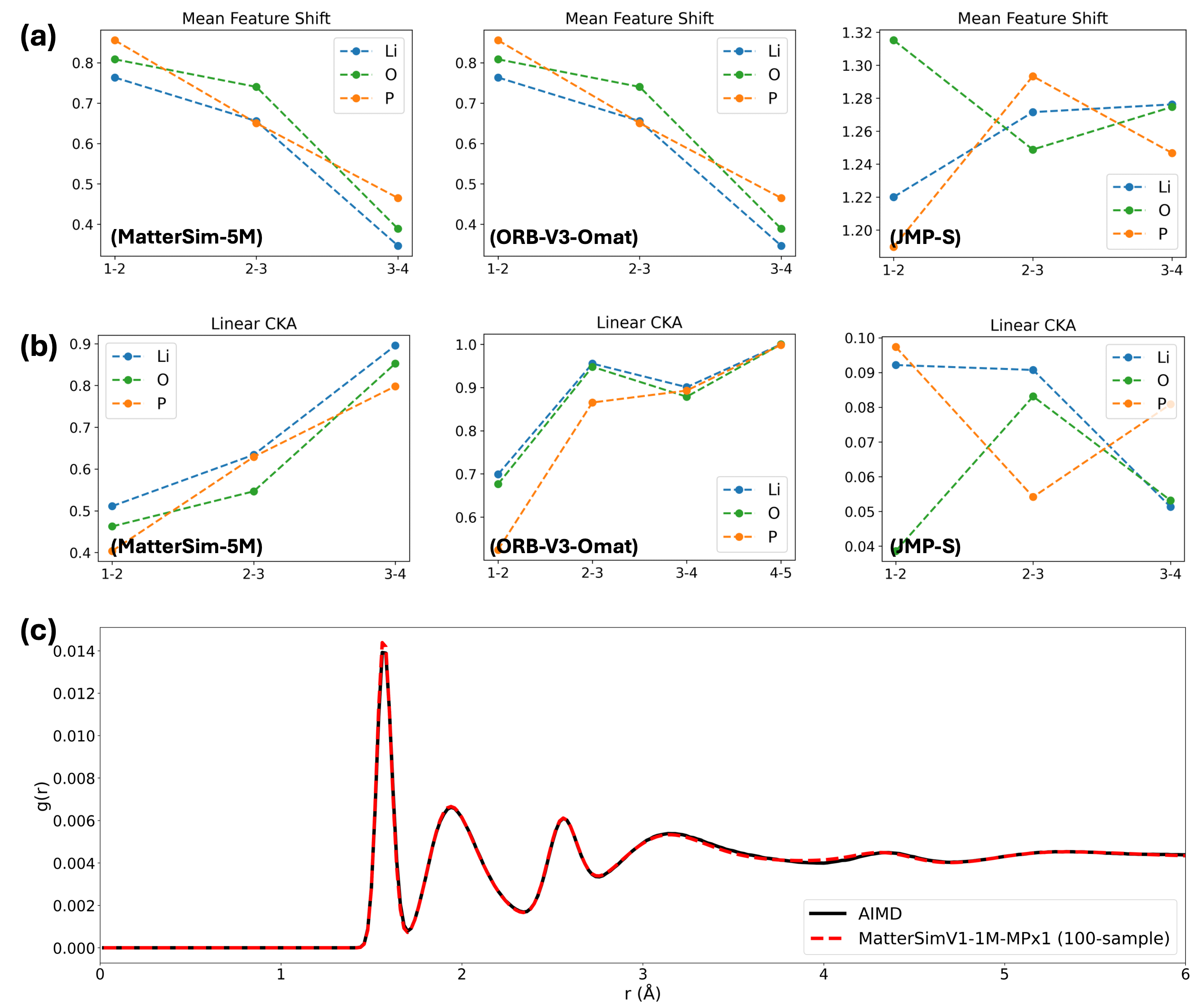}
  \caption{\textbf{(a).} Mean feature shift between successive message‑passing layers for MatterSimV1‑5M, ORB‑V3‑Omat, and JMP‑S on Li\textsubscript{3}PO\textsubscript{4}. Smaller values indicate weaker updates and greater feature stability. \textbf{(b).} Linear CKA between successive message‑passing layers for the same models. Values approaching 1.0 indicate higher similarity of node features across layers. \textbf{(c).} Radial distribution functions (RDF) of the quenched Li\textsubscript{3}PO\textsubscript{4} structure containing 421\,824 atoms. The red RDF curve was obtained from an MD trajectory driven by MatterSimV1-1M-MP$\times$1 model trained on the 100-sample dataset, whereas the black curve corresponds to the AIMD results~\cite{musaelian2023learning}.}
  \label{fig:li3po4-rdf}
  \vspace{-0.4cm}
\end{figure}

To further understand why pruning works, we analyzed the per-layer embeddings of the AFMs. We uniformly sampled 100 Li\textsubscript{3}PO\textsubscript{4} structures from the 10k‑sample dataset and collected the node embeddings produced after each successive message‑passing layer for MatterSim, ORB, and JMP. On these per‑layer embeddings we computed mean feature shift and centered kernel alignment (CKA) between every layer $i$ and layer $i+1$ as shown in Figure~\ref{fig:li3po4-rdf}. The results for MatterSim and ORB exhibit a clear depth‑dependent attenuation of feature updates: successive layers induce progressively smaller shifts and reduced diversity, indicating gradual convergence of the node representations, which is a sign of over-smoothing. Similar trends are not as pronounced for the JMP model, but the results in Table~\ref{table:li3po4-mae} indicate that the same pruning strategy is still applicable to JMP.

The pruned models substantially reduce parameter count and architectural complexity; when coupled with multi‑GPU parallel acceleration this removes prior system‑size bottlenecks and enables large-scale simulations. We evaluated the MD throughput achievable from several pruned variants under single and multi‑GPU execution on Li\textsubscript{3}PO\textsubscript{4} systems of varying sizes as shown in Table~\ref{table:li3po4-mdtime}. We observed that pruning the models led to a substantial speedup compared to their unpruned counterparts.
On top of this, multi-GPU parallelization can provide further acceleration for MD simulations, yielding more than 60\% speedup for MatterSimV1-1M and over 50\% for MACE-M-Omat.

Furthermore, Table 2 highlights another advantage of our method: since predictions on the partitioned subgraphs are completely independent, the approach can also process the subgraphs sequentially on a single GPU. This feature is particularly useful for large-scale simulations in resource-constrained  situations. Here we show that simulations of at least 5 million atoms can be performed even on a single GPU using method, far exceeding the size constraints of currently available GNN-based MLIPs (by comparison, Allegro can simulate between 200,000 to 500,000 atoms on a single 80GB A100 GPU, while larger GNNs such as MACE struggle beyond 100,000 atoms~\cite{fuchs2025chemtrain}). 

\begin{table}[H]
  \centering
  \fontsize{8pt}{11pt}\selectfont
  \begin{threeparttable}
    \caption{\textbf{MD step time (µs/(atom$\times$step)) of various AFMs on Li\textsubscript{3}PO\textsubscript{4} structures of different sizes and GPU counts (NVIDIA A100-40GB PCIe)}}
    \label{table:li3po4-mdtime}
    \begin{tabularx}{0.95\linewidth}{
        l
        l|
        >{\centering\arraybackslash}X|
        >{\centering\arraybackslash}X|
        *{3}{>{\centering\arraybackslash}X}|
        >{\centering\arraybackslash}X|
        >{\centering\arraybackslash}X
    }
      \toprule
      & & \makecell{\textbf{192}\\\textbf{atoms}}
      & \makecell{\textbf{41\,472}\\\textbf{atoms}}
      & \multicolumn{3}{c|}{\textbf{421\,824 atoms}}
      & \makecell{\textbf{943\,296}\\\textbf{atoms}}
      & \makecell{\textbf{5\,184\,000}\\\textbf{atoms}} \\
      \cmidrule(lr){3-9}
      & & \textbf{GPU$\times$1} & \textbf{GPU$\times$1} & \textbf{GPU$\times$1} & \textbf{GPU$\times$2} & \textbf{GPU$\times$4} & \textbf{GPU$\times$1} & \textbf{GPU$\times$1} \\
      \midrule
      \multirow{2}{*}{\textbf{MatterSimV1-1M}}
        & MP$\times$1 & 245.95 & 102.86 & 58.63 & 33.18 & 23.22 & 53.83 & 68.65 \\
        & MP$\times$3 & 374.34 & 374.44 & 408.00 & 204.08 & 118.12 & $\times$ & $\times$ \\
      \midrule
      \multirow{2}{*}{\textbf{ORB-V3-Omat}}
        & MP$\times$1 & 212.40 & 94.41  & 77.10  & 65.40  & 60.67 & 175.61 & 1161.92 \\
        & MP$\times$5 & 448.36 & $\times$ & $\times$ & $\times$ & $\times$ & $\times$ & $\times$ \\
      \midrule
      \multirow{2}{*}{\textbf{JMP-S}}
        & MP$\times$1 & 264.64 & 2937.86 & $\times$ & $\times$ & $\times$ & $\times$ & $\times$ \\
        & MP$\times$4 & 482.24 & $\times$ & $\times$ & $\times$ & $\times$ & $\times$ & $\times$ \\
      \midrule
      \multirow{2}{*}{\textbf{MACE-M-Omat}}
        & MP$\times$1 & 322.29 & 139.89 & 111.53 & 70.52  & 51.41 & 113.46 & 957.66 \\
        & MP$\times$2 & 533.12 & 1102.78 & 955.37 & 500.82 & 276.86 & $\times$ & $\times$ \\
      \bottomrule
    \end{tabularx}

    \begin{tablenotes}[flushleft]
      \footnotesize
      \item Reported values represent the average molecular dynamics step time in microseconds per atom per step. All benchmarks were conducted on NVIDIA A100-40\,GB PCIe GPUs using the ASE implementation of Langevin dynamics. The symbol “$\times$” indicates that even after partitioning the structure into more than 1000 subgraphs, the model could not fit within the 40\,GB GPU memory, suggesting that the model is unable or unsuitable to perform MD simulations at this scale.
    \end{tablenotes}
  \end{threeparttable}
\end{table}

Considering the trade‑off between accuracy and speed, we selected the MatterSimV1‑1M‑MP×1 model fine‑tuned on the 100‑sample dataset to reproduce the Li\textsubscript{3}PO\textsubscript{4} quenching experiment. The simulation employed the Langevin thermostat implemented in ASE~\cite{larsen2017atomic}, cooling the melt from 3000K to 600K over 50ps. From the resulting trajectory we computed the radial distribution function (RDF) of Li\textsubscript{3}PO\textsubscript{4} at 600K and compared it with the AIMD reference as shown in Figure~\ref{fig:li3po4-rdf}. The RDF matches the AIMD reference in both peak positions and relative amplitudes, while likewise lacking significant correlations past 5Å, indicating correct short and medium range structure without spurious ordering. This confirms that AFM pruned by our MPDP method not only maintains accuracy metrics in terms of energies and forces, but also preserves key physical observables critical for MD reliability.

\subsection{Nanoscale Segregation Phenomenon in the Mixed Silicon-silica System}

SiO\textsubscript{2} is known to lack a stable crystalline phase under ambient conditions and will readily disproportionate. 
Recent DFT‑based explorations of possible ordered phases of homogeneous SiO have shown that a mixture of crystalline Si and SiO\textsubscript{2} is enthalpically more stable than SiO ~\cite{alkaabi2014silicon}. Subsequently, Erhard et al. ~\cite{erhard2024modelling} employed an active learning workflow, sampling MD trajectories to construct a SiO\textsubscript{x} dataset and trained an ACE model, and was able to reproduce the segregation process of SiO into crystalline Si and SiO\textsubscript{2} via MD simulations. The modelling of the segregation process there involved lengthy melting and quenching simulations of SiO and is more complex than the Li\textsubscript{3}PO\textsubscript{4} quench. We therefore chose to reproduce this simulation to assess the robustness of the pruned AFMs under extreme, out‑of‑distribution conditions. 

To highlight the data-efficiency and robustness of AFMs, we randomly sampled 2\% of the active learning data from ~\cite{erhard2024modelling} to construct a small training set. We then applied varying degrees of pruning to MatterSim as we did before. After fine‑tuning each variant on the 2\% subset, we evaluated them on the separate test sets provided in ~\cite{erhard2024modelling} and compared the performance with the ACE‑Complex benchmark model, which was trained on all of the available data; the results are summarized in Table~\ref{table:sio-mae}. The single‑layer variant, MatterSimV1‑1M‑MP×1, attains force‑prediction accuracy comparable to the ACE‑Complex benchmark on most test sets, whereas the two‑layer variant, MatterSimV1‑1M‑MP×2, surpasses the benchmark on all test sets. A caveat here is that the energy predictions of MatterSim fine‑tuned on the 2\% subset are less accurate than the forces. We observe the same trend in Table~\ref{table:li3po4-mae}: AFMs fine‑tuned on a small subset of the full dataset tend to exhibit larger energy deviations. We attribute this primarily to sampling‑induced bias in the training energy distribution. However, in the context of the MD simulations, we find that these slightly higher energy errors did not materially affect the trajectories.

\begin{table}[h]
  \vspace{0.3cm}
  \fontsize{8pt}{11pt}\selectfont
  \centering
  \begin{threeparttable}
      \caption{\textbf{Energy and force root mean square error (RMSE) of various pruned AFMs fine-tuned on 2\% of the Si-O dataset and their comparison with the ACE baseline, reported in units of [meV] and [meV/\r{A}]}}
      \label{table:sio-mae}
      \begin{tabularx}{0.9\linewidth}{l|*{6}{Y}}
        \toprule
          &
          \multicolumn{2}{c}{\makecell[c]{ACE-Complex}} &
          \multicolumn{2}{c}{\makecell[c]{MatterSim--V1--1M\\MP$\times$1}} &
          \multicolumn{2}{c}{\makecell[c]{MatterSim--V1--1M\\MP$\times$2}} \\
          \cmidrule(lr){2-3} \cmidrule(lr){4-5} \cmidrule(lr){6-7}
          & RMSE$_E$ & RMSE$_F$ & RMSE$_E$ & RMSE$_F$ & RMSE$_E$ & RMSE$_F$ \\
        \midrule
          SiO\textsubscript{2} crystals & 0.9 & 0.05 & 120.60 & 0.08 & 146.36 & 0.04 \\
        \midrule
          a-SiO\textsubscript{2} (CHIK-MD) & 2.2 & 0.19 & 115.60 & 0.29 & 30.80 & 0.16 \\
        \midrule
          a-SiO\textsubscript{2} (GAP-MD) & 4.6 & 0.10 & 227.45 & 0.12 & 33.76 & 0.05  \\
        \midrule
          a-SiO\textsubscript{2} (ACE-MD) & 3.2 & 0.18 & 228.32 & 0.11 & 33.42 & 0.06  \\
        \midrule
          a-SiO\textsubscript{2} surfaces & 4.7 & 0.16 & 141.63 & 0.15 & 34.51 & 0.10  \\
        \midrule
          a-Si & 51.5 & 0.26 & 417.81 & 0.24 & 386.62 & 0.20  \\
        \midrule
          a-SiO\textsubscript{x} & 38.0 & 0.43 & 214.29 & 0.42 & 262.98 & 0.32 \\
        \midrule
          high-p a-SiO\textsubscript{2} & 4.6 & 0.24 & 35.84 & 0.27 & 67.19 & 0.20 \\
        \bottomrule
      \end{tabularx}
    
      \begin{tablenotes}[flushleft]
          \footnotesize
          \item All AFMs reported in this table are fine-tuned with 2\% of the entire dataset. The ACE-Complex model and its RMSE results are taken from the best-performing model reported in ~\cite{erhard2024modelling}.
      \end{tablenotes}
  \end{threeparttable}
  \vspace{0.3cm}
\end{table}

\begin{figure}[H]
  \vspace{-0.4cm}
  \centering
  \includegraphics[width=0.9\linewidth]{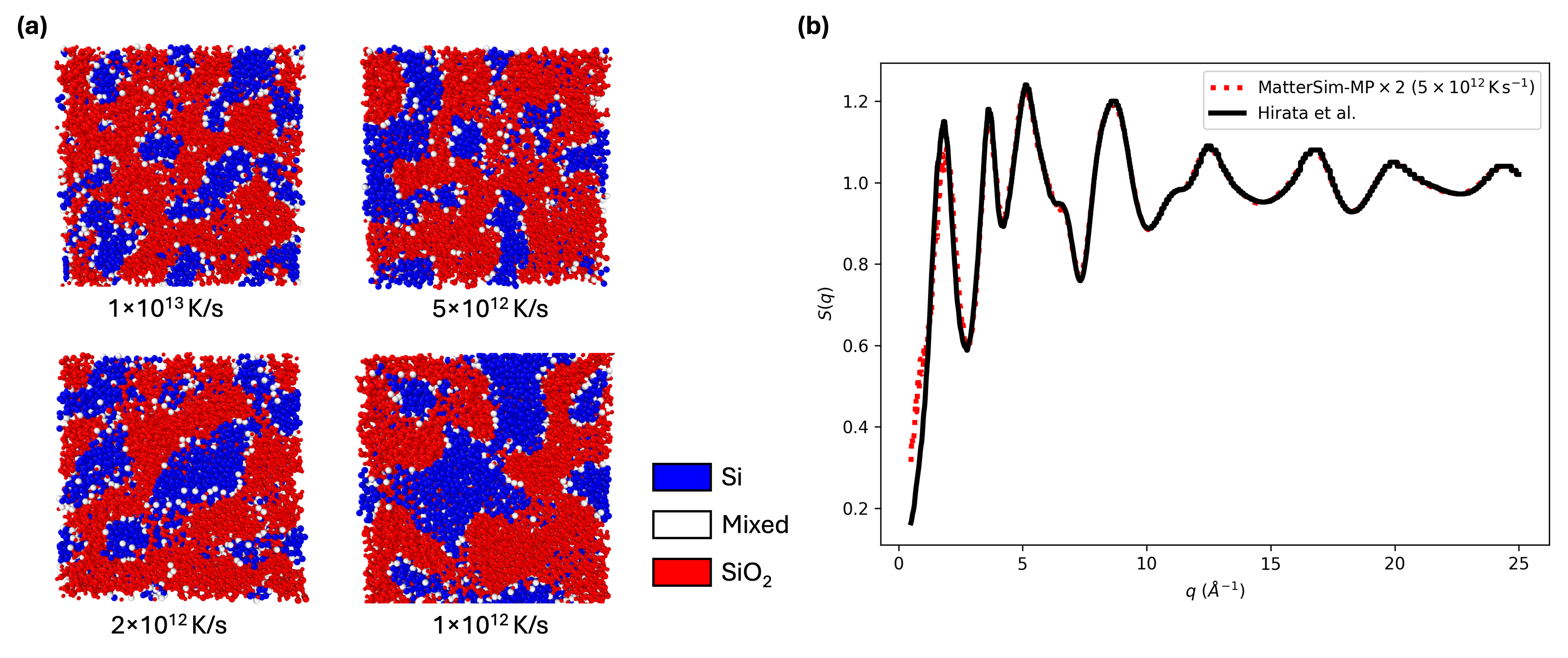}
  \caption{\textbf{(a).} Visualization of SiO structures quenched with cooling rates ranging from $10^{12}$ K/s to $10^{13}$ K/s (top left to bottom right: $1\times 10^{13}$, $5\times 10^{12}$, $2\times 10^{12}$, $1\times 10^{12}$ K/s). \textbf{(b).} Structure factor $S(q)$ of room‑temperature SiO obtained from the simulation quenched at $5\times 10^{12}$K/s, compared with the experimental reference~\cite{hirata2016atomic}. The structure factor was computed using code provided in ~\cite{erhard2024modelling}}
  \label{fig:sio-segregation}
  \vspace{-0.4cm}
\end{figure}

For the segregation simulations, we chose the MatterSimV1‑1M‑MP×1 model and performed MD simulations with ASE. A 43\,904‑atom Si/SiO\textsubscript{2} mixture was first heated in an NVT ensemble to 4000\,K and held for 10\,ps, then equilibrated for 100\,ps at 2300\,K in an NPT ensemble at zero external pressure to ensure complete melting and mixing. The system was subsequently quenched to 300\,K under zero pressure NPT using four cooling rates ranging from $10^{12}$ K/s to $10^{13}$ K/s, followed by a 10\,ps NPT hold at ambient temperature. The resulting structures are visualized in Figure \ref{fig:sio-segregation}a.  The simulations reproduce the nanoscale segregation in which distinct a-Si and a-SiO\textsubscript{2} regions appear and the average Si-rich grain size increases systematically with decreasing quench rates. We further calculated the structure factor $S(q)$ of the room‑temperature configuration generated at a quench rate of $5\times 10^{13}$ K/ and compared it with the corresponding experimental data as shown in Figure \ref{fig:sio-segregation}b. These results show the pruned AFMs can fully reproduce the observed phenomena even with 2\% of the training data.

\subsection{Metal–Support Interaction and Suboxide Encapsulation of Pt Nanocluster on TiO\textsubscript{2} Surface}

Strong metal–support interaction (SMSI) denotes the formation, under reducing or oxidizing conditions, of ultrathin overlayers of substoichiometric oxide species that migrate from an oxide support onto the supported nanoparticles, producing a coating of one to several atomic layers that partially or extensively covers the metal. This phenomena has important implications in catalysis, as SMSI can significantly alter the catalytic properties and stability of the supported nanoparticle catalysts ~\cite{tauster1981strong, liu2019ultrastable, gao2020encapsulated, polo2021situ}. Modeling SMSI at the atomic level can provide key insights into the mechanisms and characteristics of this process, providing guidance into preventing or controlling this behavior to enable more durable and active catalysts.

\begin{figure}[H]
  \vspace{-0.4cm}
  \centering
  \includegraphics[width=0.9\textwidth]{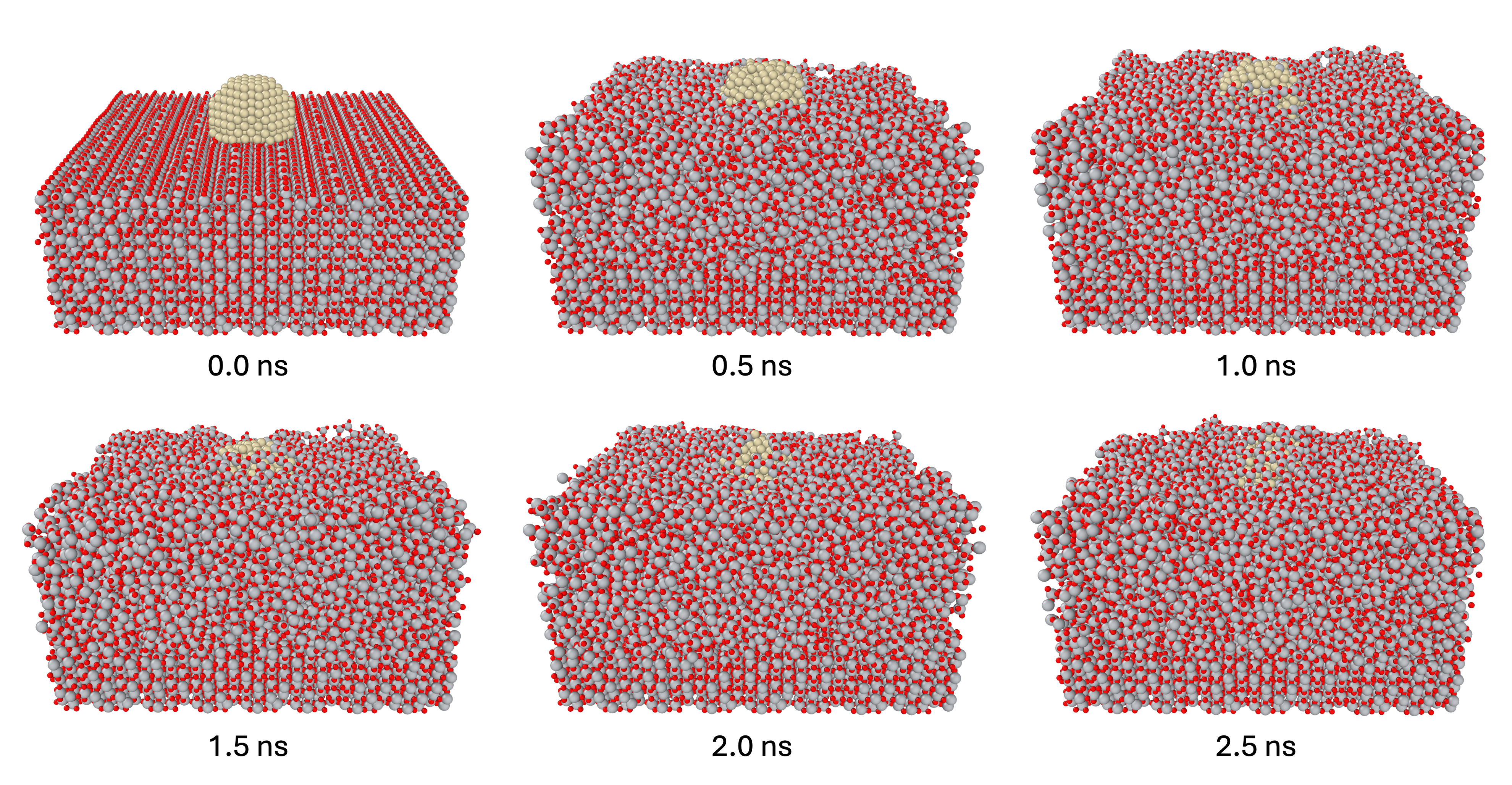}
  \caption{Snapshots from an MD trajectory showing the progressive encapsulation of a Pt nanocluster by Ti–O suboxide species on an oxygen‑deficient TiO\textsubscript{2} support. The simulation was driven by the MatterSimV1‑1M‑MP×1 model using ASE Langevin dynamics. After 2.5\,ns, the Pt cluster is fully covered.}
  \label{fig:pt-tio2}
  \vspace{-0.4cm}
\end{figure}

To assess whether our approach can reliably model these heterogeneous morphologies—slabs, nanoclusters, and their interfaces, we performed a simulation study of the SMSI process of a Pt nanocluster on a TiO\textsubscript{2} support under reducing conditions. We use a training dataset generated with DFT calculations containing smaller Pt nanoclusters of several diameters on TiO\textsubscript{x} (110) slabs and running ab-initio MD between 300K and 2000K, collecting 16\,200 configurations in total. We then fine-tuned a MatterSimV1-1M-MP$\times$1 model on these data. For the simulation, we constructed a system comprising a 4\,213-atom Pt cluster atop a 38\,400-atom TiO\textsubscript{2} slab, representing Pt nanoparticles at experimentally observed diameters. To emulate the reducing environment characteristic of catalytic operations, we uniformly removed 6.25\% of the oxygen atoms from the TiO\textsubscript{2} support, thereby introducing oxygen vacancies. We carried out an NVT simulation at an elevated temperature of 1500\,K with a 2\,fs time step to then observe the SMSI process. In Figure~\ref{fig:pt-tio2} we visualized some snapshots of the trajectory. After 2.5\,ns, we can observe the Pt cluster to be fully encapsulated by Ti–O suboxide species, which is consistent with experimental observations of SMSI-induced encapsulation. These initial results demonstrate the promise large-scale simulations with AFMs can provide towards exploring these complex phenomena for the first time, though additional work is needed to rigorously validate these simulation results with additional experimental studies. 

\section{Conclusion}

In this work, we introduced a general acceleration strategy and framework for AFMs that combines our proposed message-passing depth pruning with graph-partitioned multi-GPU parallelism. This strategy directly addresses two key obstacles to deploying AFMs in realistic molecular dynamics simulations: slow inference speeds and high memory demands. By simplifying AFM backbones to the most informative layers and distributing inference efficiently across GPUs, our framework makes large-scale and long-time simulations practical while preserving the hallmark accuracy and data efficiency of AFMs. Our experiments on three representative systems highlight both the effectiveness and the robustness of this approach across a broad range of different applications. In Li\textsubscript{3}PO\textsubscript{4}, we showed that AFMs trained on extremely small datasets can achieve competitive accuracy, and pruning further reduced complexity without compromising fidelity. In silicon-silica system and Pt/TiO\textsubscript{2} systems, we successfully capture the segregation phenomena and the metal–support encapsulation in defective oxide supports, demonstrating simulation robustness for challenging dynamics and complex heterogeneous structures. These capabilities have been integrated within the open-source MatterTune package.

Beyond the practical benefits in accelerating simulations, our findings also provide some conceptual insights for AFM design. The embedding analyses experiment revealed clear evidence of over-smoothing in AFMs: deeper message-passing layers contribute marginally to representation quality. This likely suggests that more message-passing layers do not necessarily translate into more useful information, and tailoring model depth to the target task may yield more efficient and equally accurate architectures. Further work is warranted in exploring oversmoothing in AFMs in the future.

\section{Methods}

\subsection{Message Passing Depth Pruning}

Figure~\ref{fig:overview}a schematically illustrates our strategy for constructing lightweight models by pruning message-passing layers. Consider a generic AFM backbone that employs $M$ message-passing layers to generate node and edge embeddings. Our procedure simply truncates the backbone after the first $K$ layers and feeds the resulting node and edge features directly into the property-prediction heads. This approach is applicable to the vast majority of AFM architectures. For certain models, such as JMP, which predict properties by concatenating all intermediate states before a global read-out rather than using only the final node and edge features, we preserve dimensional consistency without modifying the read-out module by padding the missing intermediate states from $K+1$ to $M$ layer with zero tensors.

\subsection{Graph Partitioning}
Our partitioning strategy ensures that each root node preserves its $K$-hop neighborhood while enabling parallel evaluation. The partitioning procedure proceeds in two stages. 

\textbf{Stage 1.} The nodes in the input atomic graph are split into disjoint sets that together cover all atoms; the atoms in each fragment constitute the root nodes of a partition. For bulk systems or other nearly uniform densities, we implemented an extremely fast spatial grid-based method: the unit cube is divided into a uniform grid of $P$ bins (with $P$ the desired number of partitions), and each atom is assigned to a bin based on its scaled position within the unit cube.

However, for heterogeneous geometries such as nanoparticles on supports, spatial partitioning alone can produce highly unbalanced or fragmented subgraphs. In such cases we have more advanced graph-based algorithms. One such algorithm, spectral partitioning, leverages the eigenvectors of the graph Laplacian to guide cuts that minimize inter-partition edges while maintaining balanced subgraphs. This approach yields high-quality partitions, but as the graph size increases, the cost of computing eigenvectors becomes prohibitive, limiting its applicability to very large systems. Thus, we instead choose to rely on METIS ~\cite{karypis1998fast}, a multilevel partitioner that approximates spectral cuts while scaling nearly linearly in the number of nodes, yielding balanced and compact subgraphs. In section \ref{sec:comparison_graph_part_methods}, we present a comparative analysis of different graph partitioning methods. 

\textbf{Stage 2.} Each subgraph is extended by a breadth first search (BFS) whose depth is equal to the pruning parameter $K$. This expansion ensures that, after partitioning, every root node retains exactly the same $K$-hop effective neighborhood it would have in the full, unpartitioned structure. To avoid any performance penalty, the BFS expansion is implemented in Cython ~\cite{behnel2010cython}. In our simulation of Li\textsubscript{3}PO\textsubscript{4} containing 421\,824 atoms, both partitioning stages together account for less than 2\% of the wall-time per step. 

\begin{figure}
    \centering
    \includegraphics[width=0.8\linewidth]{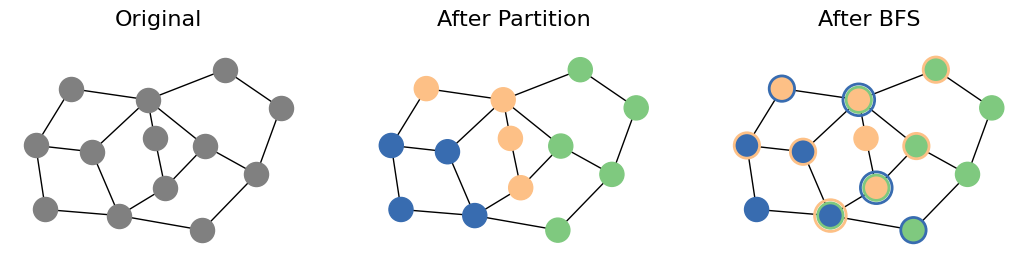}
    \caption{Example of partitioning and neighborhood expansion. Nodes are first grouped into $P=3$ disjoint partitions (colors). Boundary nodes are then expanded by a BFS of depth $K=1$, yielding overlapping regions (rings) that preserve local neighborhoods across partition boundaries}
    \label{fig:placeholder}
\end{figure}



\subsection{Global Output Integration}

After the model completes inference on each partitioned subgraph, the predicted quantities must be consistently assembled into global outputs for the full input structure, including energy, forces, and stress. For the total energy, most current AFMs decompose the system energy $E$ into a sum of per-atom contributions, i.e., $E=\sum_iE_i$. Accordingly, energy aggregation reduces to simply summing up the per-atom energy contributions of the root nodes across all subgraphs. 

For the forces, AFMs adopt two major strategies: direct prediction or energy-gradient prediction. In direct prediction, node or edge features are fed into a dedicated output head to predict forces without backpropagation. This yields computational advantages in terms of speed but does not guarantee a conservative potential energy surface. For this case, global forces can be obtained by concatenating the predicted forces of root nodes from all subgraphs. In energy-gradient prediction, forces are computed as derivatives of the energy with respect to atomic positions, i.e., $F=-\nabla E$. Since the total energy is additive over per-atom contributions, the chain rule allows us to express $\nabla E=\sum_i\nabla E_i$. Thus, it suffices to compute the gradients of root-node energies with respect to atomic positions within each subgraph and then sum these contributions across subgraphs to obtain the global force field. 

For stress predictions, a similar decomposition applies when forces are obtained via the energy-gradient approach. However, in the case of directly predicted forces, stress cannot be consistently reconstructed within our current framework.

\subsection{A Unified ASE-Compatible Multi-GPU Calculator}

We employ PyTorch Lightning’s Distributed Data Parallel (DDP) strategy ~\cite{falcon2019pytorch, li2020pytorch} to implement multi-GPU parallel inference, and based on this, we developed a unified ASE-compatible calculator ~\cite{larsen2017atomic} that supports accelerated prediction with multiple GPUs across a wide range of AFM models. This calculator, along with the model pruning and graph partitioning methods proposed in this work, has been fully integrated into the MatterTune platform ~\cite{kong2025mattertune} for public use.

\subsection{Mean Feature Shift and Linear CKA}

The mean feature shift is an average magnitude of the change in node features when moving from layer $i$ to layer $i+1$ ~\cite{dwivedi2023benchmarking, rusch2023survey}. To eliminate the effect of global translation, we normalized the feature matrices by subtracting the mean. Let $H^{(i)}\in R^{N\times d}$ and $H^{(i+1)}\in R^{N\times d}$ be the normalized node feature matrices of all $N$ nodes after successive message-passing layers. We compute mean feature shift as: 

\begin{align*}
    \text{MFS}^{(i\rightarrow i+1)}=\frac{1}{N}\lVert H^{(i+1)}-H^{(i)}\rVert_{F}
\end{align*}

where $\lVert \cdot \rVert_F$ is the Frobenius norm. 

Linear CKA measures the similarity of two whole feature spaces, insensitive to isotropic scaling or orthogonal rotation ~\cite{kornblith2019similarity}. Values that are close to 1 indicate that two feature spaces are nearly identical up to rotation/scale. To get linear CKA, we firstly center each feature matrix by $\tilde H=H-\frac{1}{N}\boldsymbol{1}\boldsymbol{1}^TH$. Then the CKA is computed as:

\begin{align*}
    \text{CKA}^{(i\rightarrow i+1)}=\frac{\lVert\tilde H^{(i)\top}\tilde H^{(i+1)}\rVert^2_F}{\lVert\tilde H^{(i)\top}\tilde H^{(i)}\rVert_F\lVert\tilde H^{(i+1)\top}\tilde H^{(i+1)}\rVert_F}
\end{align*}

\section{Acknowledgements}
The support of this work by the National Science Foundation grant CBET-2442223 is gratefully acknowledged. This research used resources of the National Energy Research Scientific Computing Center, a DOE Office of Science User Facility supported by the Office of Science of the U.S. Department of Energy under Contract No. DE-AC02-05CH11231 using NERSC award BES-ERCAP0032102.  
\section{Code Availability}

The MatterTune platform, used to perform the experiments in this paper, are available on Github at https://github.com/Fung-Lab/MatterTune.

\bibliographystyle{unsrt}
\bibliography{reference}

\clearpage
\appendix
\section*{Supplementary Information}
\addcontentsline{toc}{section}{Supplementary Information}

\setcounter{section}{0}
\renewcommand{\thesection}{S\arabic{section}}
\renewcommand{\thesubsection}{S\arabic{section}.\arabic{subsection}} 

\setcounter{figure}{0}
\setcounter{table}{0}
\setcounter{equation}{0}
\renewcommand{\thefigure}{S\arabic{figure}}
\renewcommand{\thetable}{S\arabic{table}}
\renewcommand{\theequation}{S\arabic{equation}}

\section{Comparison of Graph Partitioning Methods}
\label{sec:comparison_graph_part_methods}

\begin{figure}[H]
    \centering
    \includegraphics[width=0.8\linewidth]{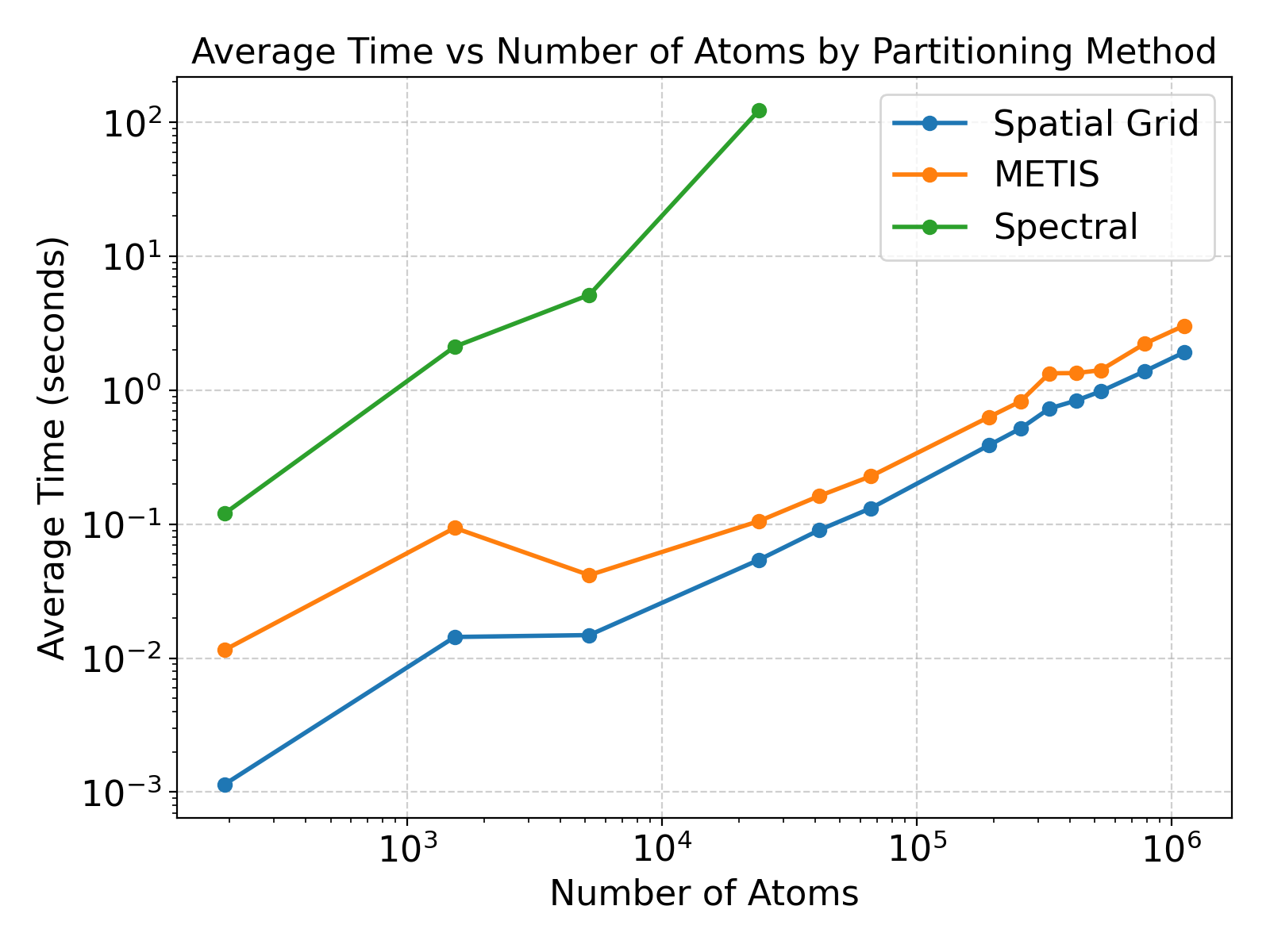}
    \caption{Runtime comparison of different graph partitioning algorithms on Li\textsubscript{3}PO\textsubscript{4} structures of varying sizes.}
    \label{fig:figs1}
\end{figure}

To assess both the runtime characteristics of different graph partition methods and their behavior across structures with distinct geometries, we performed a head-to-head evaluation. For each system size, we partitioned the atomic graph into $8$ subgraphs. Under this common setting, we recorded the wall-clock time required by three algorithms—spatial grid partitioning, METIS, and spectral partitioning (Fig. \ref{fig:figs1}). Spectral partitioning becomes prohibitively expensive once the system size exceeds several tens of thousands of atoms, whereas METIS and the spatial grid exhibit substantially flatter scaling; owing to its simplicity, the spatial grid is consistently the fastest.

\begin{figure}[H]
    \centering
    \includegraphics[width=0.8\linewidth]{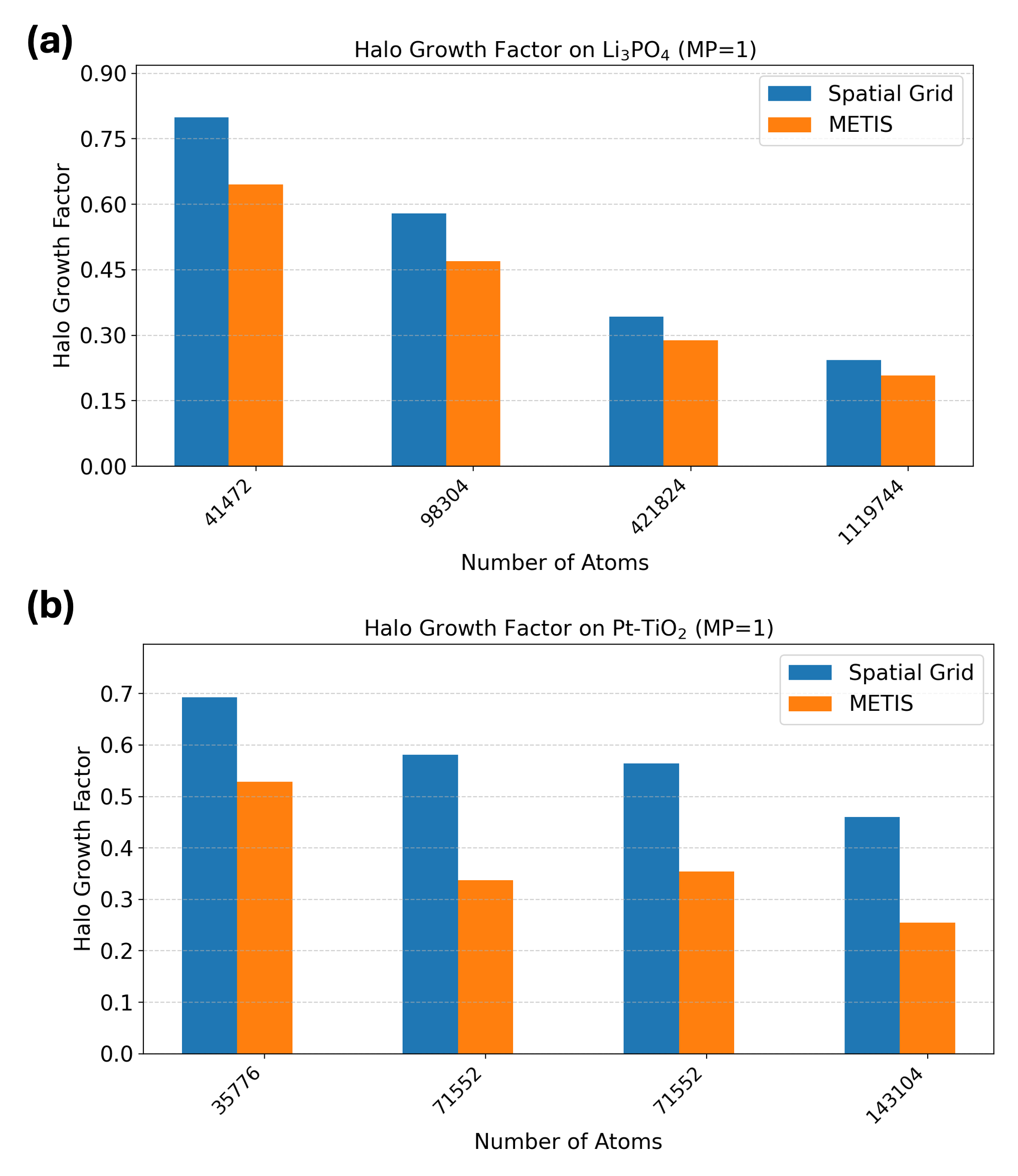}
    \caption{Comparison of halo growth factor across graph partitioning algorithms for (a) Li\textsubscript{3}PO\textsubscript{4} structures and (b) Pt-TiO\textsubscript{2} structures, evaluated over varying system sizes.}
    \label{fig:figs2}
\end{figure}

We further compared the quality of partitions produced by METIS and the spatial grid using the halo growth factor $H$ on Li\textsubscript{3}PO\textsubscript{4} and Pt-TiO\textsubscript{2} systems of various sizes (Fig \ref{fig:figs2}). We define the halo growth factor $H$ as the ratio of the atoms introduced by halo expansion to the number of atoms in the original subgraph. Smaller $H$ indicates less overhead from halo expansion and thus higher-quality partition boundaries. On the homogeneous bulk Li\textsubscript{3}PO\textsubscript{4} structures, METIS holds only a slight advantage over the spatial grid. In contrast, for the heterogeneous Pt-TiO\textsubscript{2}, which are spherical particles placed on planar slabs, METIS achieves a noticeably lower halo growth factor, indicating a clearer advantage. 

\end{document}